\pdfoutput=1
\documentclass[11pt]{article}

\usepackage[]{ACL2023}

\usepackage{times}
\usepackage{latexsym}

\usepackage[T1]{fontenc}
\usepackage[utf8]{inputenc}

\usepackage{microtype}

\usepackage{inconsolata}

\usepackage{enumitem}
\usepackage{caption}
\usepackage{subcaption}
\usepackage{multirow}

\usepackage{amsmath}
\usepackage[capitalise]{cleveref}
\usepackage{graphicx}

\usepackage{tabularx}
\usepackage{graphicx}
\usepackage{booktabs}       % professional-quality tables
\usepackage{amsfonts}       % blackboard math symbols
\usepackage{nicefrac}       % compact symbols for 1/2, etc.
\usepackage{microtype}      % microtypography
\usepackage{xcolor}  

\usepackage{url}
\usepackage{hyperref}

\newcommand{\MD}{{D}}
\newcommand{\MC}{{C}}

\newcommand{\MG}{{G}}

\title{Generation-Augmented Query Expansion For Code Retrieval}

\author{
  Dong Li\textsuperscript{1}\thanks{ \ \ Work was done when Dong Li were interning at Microsoft.}  , 
  Yelong Shen\textsuperscript{2}, 
  Ruoming Jin\textsuperscript{1},
  \textbf{Yi Mao}\textsuperscript{2} , \\
  \textbf{Kuan Wang}\textsuperscript{3}, 
  and \textbf{Weizhu Chen}\textsuperscript{2}\\
  \textsuperscript{1}Kent State University \\
  \textsuperscript{2}Microsoft \\
  \textsuperscript{3}Georgia Institute of Technology\\
  \texttt{\{dli12, rjin1\}@kent.edu} \ \ \  \texttt{\{yelong.shen, maoyi, wzchen\}@microsoft.com} \ \ \ \texttt{kuanwang@gatech.edu}\\
}

\begin{document}
\maketitle
\begin{abstract}
Pre-trained language models have achieved promising success in code retrieval tasks, where a natural language documentation query is given to find the most relevant existing code snippet. However, existing models focus only on optimizing the documentation code pairs by embedding them into latent space, without the association of external knowledge. In this paper, we propose a generation-augmented query expansion framework. Inspired by the human retrieval process - sketching an answer before searching, in this work, we utilize the powerful code generation model to benefit the code retrieval task. Specifically, we demonstrate that rather than merely retrieving the target code snippet according to the documentation query, it would be helpful to augment the documentation query with its generation counterpart - generated code snippets from the code generation model. To the best of our knowledge, this is the first attempt that leverages the code generation model to enhance the code retrieval task. We achieve new state-of-the-art results on the CodeSearchNet benchmark and surpass the baselines significantly. 
\end{abstract}

\section{Introduction}
%what is the problem you study, and the importance of the problem  
%state-of-the-art research 
\begin{figure}
    \centering
    \includegraphics[width=\linewidth]{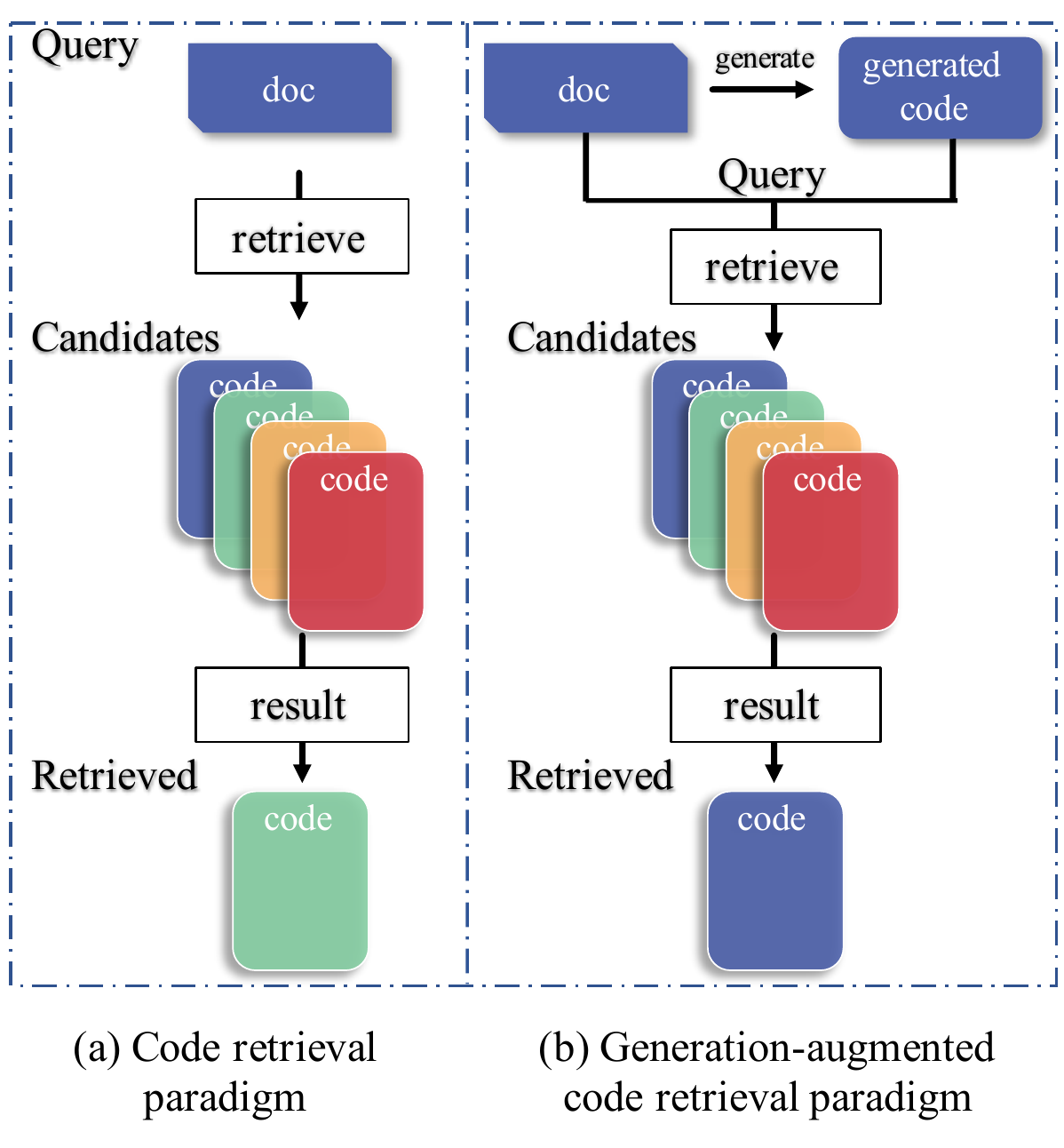}
    \caption{Illustration of (a) code retrieval paradigm and (b) generation-augmented code retrieval paradigm.}
    \label{fig:intro_compare}
\end{figure}

\begin{figure*}
    \centering
    \includegraphics[width=\linewidth]{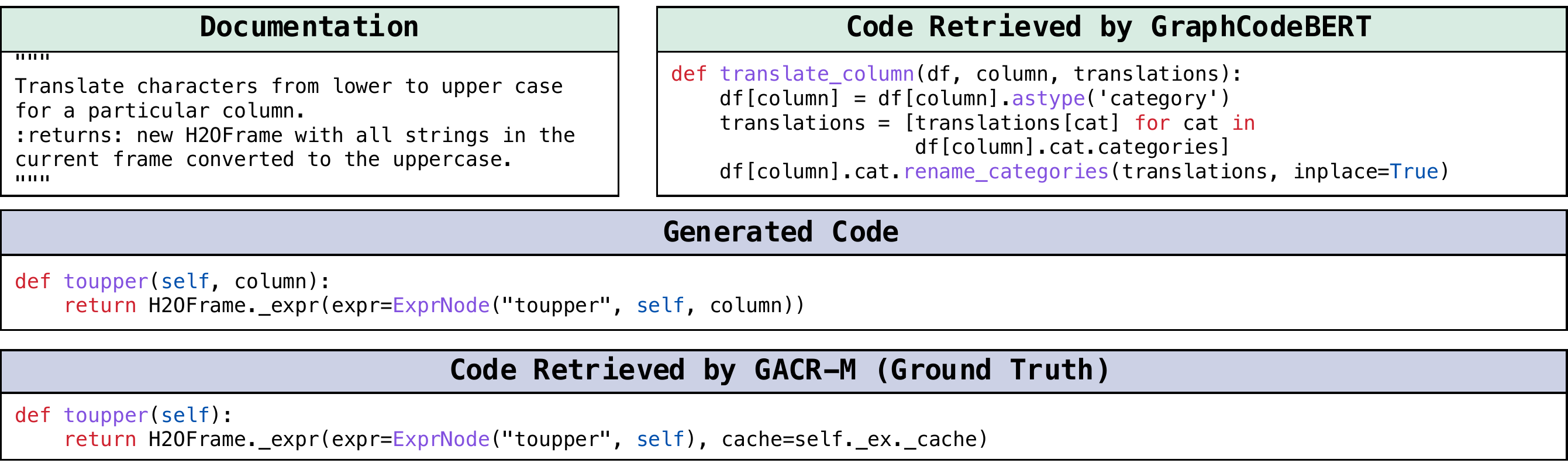}
    \caption{An example that generated codes help documentation query to find the ground truth. Given an NL presented "Documentation", the primary goal is to retrieve the "Ground Truth" code snippet from a candidate pool, while the "Code Retrieved by GraphCodeBERT" is the actual one that the retrieval model (with doc query) \cite{guo2021graphcodebert} finds. After incorporating "Generated Code", the retrieval model proposed in this paper is able to retrieve the "Ground Truth" Code as its first choice.}
    % In this example, the ("Documentation", "GroundTruth Code") pair is a real data point in CodeSearchNet dataset \cite{codesearchnet} from \href{https://github.com/h2oai/h2o-3/blob/dd62aaa1e7f680a8b16ee14bc66b0fb5195c2ad8/h2o-py/h2o/frame.py\#L2924-L2930}{github}, same as "Retrieved Code" from \href{https://github.com/okfn-brasil/serenata-toolbox/blob/47b14725e8ed3a53fb52190a2ba5f29182a16959/serenata_toolbox/datasets/helpers.py\#L43-L52}{github}.}
    \label{fig:1434}
\end{figure*}

Benefiting from the development of transformers \citep{transformer,codebert,guo2021graphcodebert} as well as pre-training techniques \citep{CodeRetriever_Li2022}, there has been a great amount of progress in code-related tasks, including code generation \citep{gpt-3}, code search \citep{codesearchnet}, code auto-completion \citep{lu2022reacc}, etc. Among these, \textit{Code search} or \textit{code retrieval} \citep{codesearchnet} is an essential problem in software engineering, which enables the efficient finding and reuse of existing code snippets, boosting developers' productivity. In general, it aims to retrieve function-level code snippets given a natural language documentation query. Recently, CodeBERT \citep{codebert} has achieved great performance in code-related tasks by pre-training a bimodal model to learn a general-purpose representation for both programming language (PL) and natural language (NL). By considering the inherent code structure - data flow, GraphCodeBERT ~\cite{guo2021graphcodebert} further boost the performance on downstream tasks. Meanwhile, CodeRetriever ~\cite{CodeRetriever_Li2022} adopts the contrastive-training approach to learn the semantic presentation of code on a large-scale pretraining corpus.

%inefficiency; drawback... the limitation 
Despite of the success of all these code-specific frameworks \citep{codebert,guo2021graphcodebert,CodeRetriever_Li2022,guo-etal-2022-unixcoder,codesearchnet}, they only focus on learning the information within the code-documentation context (either by optimizing specific downstream tasks or by contrastively learning general-purpose representations), without considering external knowledge, which limits the expressiveness of the representations.
Specifically for code retrieval tasks, due to the intrinsic difference between the NL and PL, the current end-to-end code search models are insufficient to retrieve the most semantically similar code snippet. Since documentation queries and code snippets are typically embedded into the same latent space, the model would try to find an "efficient" shortcut - matching two vectors once some keywords are triggered. For example, in \Cref{fig:1434}, "Documentation" on the left-top is a natural language, an end-to-end model \citep{guo2021graphcodebert} would find the "Retrieved Code by GraphCodeBERT" in the right-top whereas they share a few common keywords. This misleading result testifies to the drawbacks of these models.

 % your idea and why it is novel!!  
 To tackle this issue, we propose GACR, \textbf{G}eneration-\textbf{A}ugmented \textbf{C}ode \textbf{R}etrieval, a two stage framework for code retrieval task, illustrated in \Cref{fig:intro_compare}. First, the natural language (NL) text documentation is used to generate code snippets by a code generation model (for example, GPT-3). Then the generated code augmented NL documentation would serve as an expanded query for retrieval. The benefit of GACR is that it could leverage extra domain knowledge specifically generating code snippets according to a natural language-described documentation prompt. This would bridge the gap between the NL and PL which has been overlooked in current end-to-end code retrieval models. Considering the same example in \Cref{fig:1434}, by leveraging the power of the generation model, the "Documentation" could be used to generate a code snippet - the "Generated Code", which is much more semantically close to the "Ground Truth". And thus, our proposed model - GACR is able to find the desired code.
 
% new idea, challenges and how to solve this challenges
 Though there are some work \citep{Retrieval_Aug_Code_Parvez2108,lu2022reacc} that leverage retrieval model to help generation, enhancing code retrieval tasks with generation model lacks attention. The main challenges are how to generate code snippets with good quality and fusion them with documentation queries. Thanks to the promising achievement of GPT \citep{gpt-3}, we select the Codex, an ad-hoc fine-tuned generation model on publicly available code datasets. As to the utilization, we design a dual representation attention paradigm that allows learning information from NL documentation and PL generated code snippet respectively, and mutually. The expansion queries fuse the contents from NL and PL, leading to informative and expressive representations.
 
%Your detailed contribution: 
We evaluate the performance of the GACR on CodeSearchNet benchmark \cite{codesearchnet} with 6 programming languages. The empirical results show that the proposed \textit{generation-augmented code retrieval} models achieve significant improvements compared to the baseline models \cite{guo2021graphcodebert, CodeBERT_Feng2020}. To this end, we summarize our contributions as follows:
\begin{itemize}[leftmargin=0.15in]
\setlength\itemsep{-0.5em}
    \item We propose a generation-augmented framework for code retrieval tasks. To the best of our knowledge, this is the first attempt that leverages the power of the generation model to help with code retrieval tasks.
    \item We design generation-augmentation frameworks, spinning from the utilization of a single generated code snippet to multiple distinct code snippets to adapt various deployments and investigate the fusion pattern of the NL documentation and PL generated code, as well as different pre-train models.
    \item Extensive empirical experiments on the CodeSearchNet benchmark validate the superiority of the proposed augmentation paradigm.
\end{itemize}
The following of the paper is organized as: in \Cref{sec:background}, we introduce the background and formalize the notations. In \Cref{sec:model}, we talk about the details of how we incorporate the generated code for retrieval tasks. In \Cref{sec:experiment}, we conduct experiments and analysis empirical results. In \Cref{sec:related}, we summarize a few related works and in \Cref{sec:conclustion}, we conclude the paper. Further, in \Cref{sec:case}, we show specific cases of how the generated code snippets boost or depress the performance on retrieval tasks.

\section{Background and Notation}\label{sec:background}

\begin{figure*}[]
    \centering
    \includegraphics[width=\linewidth]{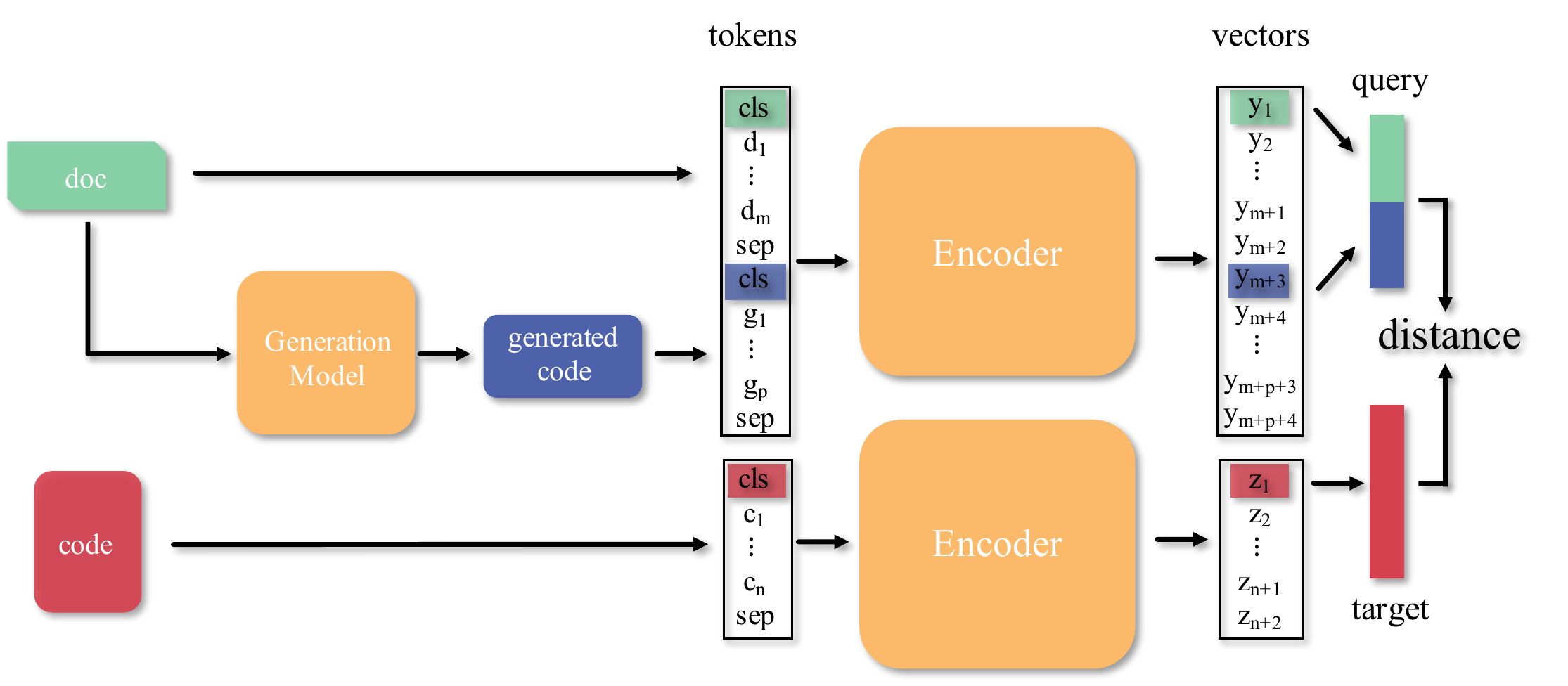}
    \caption{An illustration of generation-augmented code retrieval framework.}
    \label{fig:illustration}
\end{figure*}

\subsection{Dense Retrieval}\label{subsec:background-coderetrieval}
Encoder-based dense representation frameworks have achieved great success in retrieval tasks \citep{latentIR@shen, karpukhin-etal-2020-dense,codesearchnet,zhang2022ar2}. In short, queries and documents are mapped into the latent space, and similarity metrics are computed according to their distance (in the latent space). Specifically for code retrieval tasks, there are {documentation} - {code snippet} ($\MD,\MC$) pairs in the training sets. Denote the sequence of natural language documentation tokens as $\MD=\{d_1,d_2,\dots,d_m\}$ with length $m$ and each $d_i (i\in\{1,2,\dots,m\})$ is a token. Similarly,
the sequence of code snippet is denoted as $C=\{c_1,c_2,\dots,c_n\}$. The encoder would take the pairs as input and map them into vector representations, such that we have $V_{d}=\textbf{Encoder}(\MD)$ and $V_{c}=\textbf{Encoder}(\MC)$. The semantic similarity between document and code can be simply defined as the dot product: 
\begin{equation}\label{eq:inner_score}
    \begin{split}
        S_{score}=V^T_{d} \cdot V_{c}
    \end{split}
\end{equation}
In the training stage, the encoder is optimized to maximize the scores in \Cref{eq:inner_score} for related {documentation} - {code snippet} pairs which is analogous to minimizing the distance between them in latent space. In the inference stage, the similarity scores between the query vector (obtained by feeding $\MD$ into the encoder) and all the candidate code vectors would be computed according to \Cref{eq:inner_score}. The retrieved code snippets are defined as the ones that have large scores.

% the encoder would transform each into an individual latent vector or so-called dense representation. 

% \subsection{Formulation}\label{subsec:formulation}
% wewewe
% We denote the sequence of natural language documentation tokens as $\mathcal{D}=\{d_1,d_2,\dots,d_m\}$ with length $m$ and each $d_i (i\in\{1,2,\dots,m\})$ is a token. Similarly,
% the sequence of code snippet (in candidate pools to be retrieved) is denoted as $C=\{c_1,c_2,\dots,c_n\}$ and 
%  the sequence of generated code snippet is noted as $G=\{g_1,g_2,\dots,g_p\}$. Generated code snippet $G$ is obtained by feeding documentation $D$ into some well-trained generation model say $GEN$, $G=GEN(D)$. In stead of feeding merely documentation $D$ into the encoder to obtain a query vector $V_{query}=Encoder(D)$, we also utilize generated code: $V_{query} = Encoder(D,G)$. Meanwhile, all codes in candidate base would have their own vector representations $V_{target}=Encoder(G)$. Similarities between query and target vectors would further determine the selection of retrieval.

\subsection{Prompt-based Code Generation}\label{subsec:background-generation}
Large pre-trained language models have demonstrated an awesome power to generate code \citep{codex,gpt-3,gpt-j,Nijkamp2022CG,alphacode}. Prompt-based code generation or completion model \citep{codex} would take the natural language prompt as the input and output the corresponding code snippet. In this paper, we mainly study Codex \citep{codex} model, a GPT \citep{gpt-3} based language model that is fine-tuned on publicly available code collected from GitHub and particularly to generate functionally code snippets given NL documentation strings. In our setting, the documentation $\MD$ is prompt. After feeding it into the generation model (Codex), we would obtain the generated code snippet: $\MG=\textbf{Gen}(\MD)$. Presenting in a sequence way: $\MG=\{g_1,g_2,\dots,g_p\}$ with length $p$, where $g_i$ is a code token.

\section{Methodology}\label{sec:model}

%Instead of treating NL documentation alone as a query to retrieve the corresponding code snippets, in this section, 
We propose to enhance the documentation query with its generation counterpart for code retrieval tasks in this section. At a high level, we first feed the given query $\MD$ into a code generation model powered by Codex, which serves as a prompt, and generates a code snippet $\MG$ accordingly.
% $G=GEN(D)$. We denote the sequence of generated code snippet is noted as $G=\{g_1,g_2,\dots,g_p\}$ with length $p$.
Afterward, the documentation along with its generated auxiliary code would be treated as the query vector jointly:
$V_{query} = \textbf{Encoder}(D,G)$, used to retrieval the most correlated code snippets from the candidate pool.

\subsection{Query Augmentation with Single Generated Code}\label{subsec:single_code}

One strategy is to append the generated auxiliary code tokens to the end of the documentation query and then feed it into the encoder model, obtaining the vector representation. It is worth being aware that the generated code and original documentation come from distinct semantic domains - natural language (NL) and programming language (PL), leading to a drawback if we take one single representation for them. Thus, we design a dual representation attention paradigm - two semantic input sequences (documentation and generated code) could possess their own vector representations respectively while the attention mechanism would elegantly ensure learning the information mutually.  

The overall model architecture is shown in \Cref{fig:illustration}. Given a sequence of documentation tokens $D$. We feed the documentation sequence $D$ into a well-trained code generation model - Codex (aka GPT-3) and obtain the output - generated code token sequence $G$. Further, two sequences are concatenated together, by some special tokens, as a single input sequence: $X =\{D, G\}= \{[CLS],d_1,\cdots,d_m,[SEP],$ $[CLS], g_1,\cdots,g_p,[SEP]\}$ and the length of which is  $m + p + 4$. 

The input sequence $X$, consisting of fusion tokens from documentation and generated code, shall further be converted as the vector representation $Y$. We denote the transformer operation as $T$. After feeding the input $X$ into the $K$ layers multi-head self-attention transformer model \citep{transformer,guo2021graphcodebert}, we shall obtain a vector sequence $Y = \{y_1,y_2\cdots,y_{m+p+4}\}$, with the same length as the input sequence:
\begin{equation}
    \begin{split}
Y = T_K\circ T_{K-1}\cdots T_1\circ H^0
    \end{split}
\end{equation}
 The vectors at index $1$ and $m+3$ of $Y$, corresponding to the $CLS$ tokens of documentation and generated code, would be extracted and concatenated as the final query vector $V_{query} = [y_1, y_{m+3}]$. Correspondingly, the target vector is replicated for dimension: $V_{target} = [z_1,z_1]$.

\subsection{Augmentation with Multi-Generated Codes}\label{subsec:multi-code}

In \Cref{subsec:single_code}, we proposed the framework that allows the incorporation of a single generated code snippet into the documentation query. Here we extend the framework and include multiple distinct generated code snippets to further generalize and boost the model. In short, multiple generated code snippets are integrated before they are fed into the Encoder. To enable a diverse expansion, we limit the length of each code snippet. Given $k$ unique generated code snippets, the input token sequence can be formularized as $X = \{D, G_1,\cdots, G_k\}$, where $D$ represents the sequence of documentation tokens and $G_i, i\in \{ 1,\cdots, k\}$ is the $i$-th generated code token sequence. Similar to \Cref{subsec:single_code}, we still extract two special vectors, corresponding to the first and second $CLS$ tokens (ahead of $D$ and $G_1$), from the output vector sequence $Y$.

In this case, all generated codes are mixed together before they are fed into the encoder, that's so-called "pre-attention" - fusion before the attention. This pre-attention mechanism allowed all distinct generated code snippets to learn from each other mutually.

\begin{table*}[]
\caption{Model performance on CodeSearchNet \citep{codesearchnet} benchmark (further crafted by \citep{guo2021graphcodebert}). We highlight in bold the best model and underline the second best one in each column. We also list the relative improvements in percent w.r.t the initialization base models (reproduced ones of  GraphCodeBERT and CodeRetriever, respectively).}
\label{tab:main_table_1}
\centering
\resizebox{\linewidth}{!}{%
\begin{tabular}{cccccccc}
\cmidrule(lr){1-8} 
\textbf{Model	$\backslash$ Language}       & \textbf{Ruby}  & \textbf{Javascript} & \textbf{Go}    & \textbf{Python} & \textbf{Java}  & \textbf{Php}   & \textbf{Overall} \\
\cmidrule(lr){1-8} 
RoBERTa              & 0.587          & 0.517               & 0.850          & 0.587           & 0.599          & 0.560          & 0.617            \\
RoBERTa (Code)       & 0.628          & 0.562               & 0.859          & 0.610           & 0.620          & 0.579          & 0.643            \\
CodeBERT             & 0.679          & 0.620               & 0.882          & 0.672           & 0.676          & 0.628          & 0.693            \\
UniXcoder	&0.740	&0.684	&0.915	&0.720	&0.726	&0.676	& 0.744\\
\cmidrule(lr){1-8} 
% \multicolumn{1}{l}{} & \multicolumn{7}{c}{initialized by GraphCodeBERT}  \\
% \cmidrule(lr){2-8} 
\multicolumn{8}{l}{\textit{initialized by GraphCodeBERT pre-trained base}}  \\
GraphCodeBERT (report)        & 0.703          & 0.644               & 0.897          & 0.692           & 0.691          & 0.649          & 0.713            \\
GraphCodeBERT (reproduce)        & 0.703          & 0.644               & 0.890          & 0.693          & 0.695          & 0.646          & 0.712            \\

GACR-S       & 0.772 (9.7\%)          & 0.747 (16.0\%)              & 0.893 (0.4\%)          & 0.768 (10.8\%)          &{0.726 (4.4\%)} & 0.813 (25.8\%)          & 0.786  (10.4\%)          \\
GACR-M         & \underline{0.798} (13.5\%) & \underline{0.767} (19.1\%)     & {0.898 (0.9\%)} & 0.766 (10.6\%)          & 0.686  (-1.2\%)        & \underline{0.825} (27.8\%) & {0.790 (11.0\%)}    \\
\cmidrule(lr){1-8} 
% \multicolumn{1}{l}{} & \multicolumn{7}{c}{initialized by CodeRetriever}   \\
% \cmidrule(lr){2-8} 
\multicolumn{8}{l}{\textit{initialized by CodeRetriever pre-trained base}}   \\
CodeRetriever (report)       & 0.753          & 0.695               & \textbf{0.916} & 0.733           & \underline{0.740}          & 0.682          & 0.753            \\
CodeRetriever (reproduce)       & 0.727          & 0.668               & {0.901} & 0.704           & 0.707          & 0.656          & 0.727            \\

GACR-S            & 0.789 (8.5\%)         & 0.762 (14.1\%)              & 0.901 (-0.1\%)          & \underline{0.780} (10.8\%)            & \textbf{0.741} (4.8\%) & 0.814 (24.0\%)          & \underline{0.798} (9.7\%)           \\
GACR-M            & \textbf{0.808} (11.1\%) & \textbf{0.778} (16.5\%)      & \underline{0.902} (0.1\%)         & \textbf{0.782} (11.0\%)  & 0.727 (2.9\%)         & \textbf{0.828} (26.3\%) & \textbf{0.804} (10.6\%)  \\
\cmidrule(lr){1-8} 
\end{tabular}
}
\end{table*}

\subsection{Optimization and Inference}
In the training stage (for each batch $B$), we are given $|B|$ related/positive documentation-code pairs $(D^b, C^b)$, $b\in \{1,2,\cdots, |B|\}$. Either with single-generated code augmentation or multi one, we shall have vectors pairs $(V^b_{query}, V^b_{target})$ correspondingly (see \Cref{fig:illustration}). Similar to \citep{codesearchnet,guo2021graphcodebert,CodeRetriever_Li2022}, an in-batch optimization is utilized:
\begin{equation}
    \begin{split}
        \mathcal{L}=-\frac{1}{B}\sum\limits_{b\in B}{\frac{ \exp{\Big( (V^b_{query})^T\cdot V^b_{target} } \Big)}{ \sum\limits_{j\in B}{\exp{\Big( (V^b_{query})^T\cdot V^j_{target} } \Big)}}}
    \end{split}
\end{equation}

As to the inference stage, the similarity score can be calculated by \Cref{eq:inner_score} for each query $D$ w.r.t all the candidate code snippets. And the desired retrieval code snippets can be obtained by ranking scores accordingly.

\section{Experiments}\label{sec:experiment}

% \begin{table*}[]
% \label{tab:main_table_1}
% \caption{Main Results}
% \centering
% \begin{tabular}{cccccccc}
% \cmidrule(lr){1-8} 
% \textbf{Model} & \textbf{Ruby}  & \textbf{Javascript} & \textbf{Go}    & \textbf{Python} & \textbf{Java}  & \textbf{Php}   & \textbf{Overall} \\
% \cmidrule(lr){1-8} 
% RoBERTa        & 0.587          & 0.517               & 0.85           & 0.587           & 0.599          & 0.56           & 0.617            \\
% RoBERTa (Code) & 0.628          & 0.562               & 0.859          & 0.61            & 0.62           & 0.579          & 0.643            \\
% CodeBERT       & 0.679          & 0.62                & 0.882          & 0.672           & 0.676          & 0.628          & 0.693            \\
% GraphCodeBERT  & 0.703          & 0.644               & 0.897          & 0.692           & 0.691          & 0.649          & 0.713            \\
% \cmidrule(lr){1-8} 
% Generated Code as query& 0.707          & 0.681               & 0.844          & 0.650           & 0.594          & 0.732          & 0.701            \\
% query augmented by GC        & 0.772          & 0.747               & 0.893          & 0.768           & \textbf{0.726} & 0.813          & 0.786            \\
% pre-attention multi-code          & \textbf{0.798} & \textbf{0.767}      & \textbf{0.898} & 0.766           & 0.686          & \textbf{0.825} & \textbf{0.79}    \\
% post-attention multi-code         & 0.779          & 0.755               & 0.895          & \textbf{0.772}  & 0.717          & 0.822          & \textbf{0.79}   \\
% \cmidrule(lr){1-8} 
% \end{tabular}
% \end{table*}

We compare the performance of the proposed framework - GACR with a number of state-of-the-art baseline models:
\begin{itemize}[leftmargin=0.15in]
\setlength\itemsep{-0.5em}
    \item \textbf{RoBERTa}\citep{roberta} is a robustly optimized BERT pretraining approach , while the \textbf{RoBERTa (code)} is pre-trained sorely on the code.
    \item \textbf{CodeBERT}\citep{codebert}, a bimodal pre-trained
model for PL and
NL.
    \item \textbf{GraphCodeBERT}\citep{guo2021graphcodebert}, extract the topological information from the code and build a data flow graph, achieving state-of-the-art results on several downstream tasks. 
    \item \textbf{UniXcoder}\citep{guo-etal-2022-unixcoder}, a unified  cross-modal pre-trained model enhancing the representation by utilizing information from code comment and AST.
    \item \textbf{CodeRetriever}\citep{CodeRetriever_Li2022}, learning with unimodal and bimodal two contrastive schemes and achieving the start-of-the-art in the code search task. 
\end{itemize}

% A code search task is: given the documentation or context of the code, which is natural language, retrieve the most semantically related code.
% In \cite{guo2021graphcodebert}, author extract the topological information from the code and build a data flow graph. By constructing the graphical attention mechanism, the model achieve good results on several downstream tasks.

The experiments are carried on CodeSearchNet code corpus datasets, which are initially released by \cite{codesearchnet} and further handcrafted by \cite{guo2021graphcodebert} for the code quality reason. Here, we take the same settings as \cite{guo2021graphcodebert}. Overall, all models (both baselines and proposed ones) derive the vector representations for query and candidates, then compute the dot product of them as a score for rank and retrieve. The performance is measured in terms of Mean Reciprocal Rank (MRR).

\subsection{RQ1. How does the generation-augmented framework - GACR perform compared to the baselines?}

\begin{figure*}
    \centering
    \includegraphics[width = \linewidth]{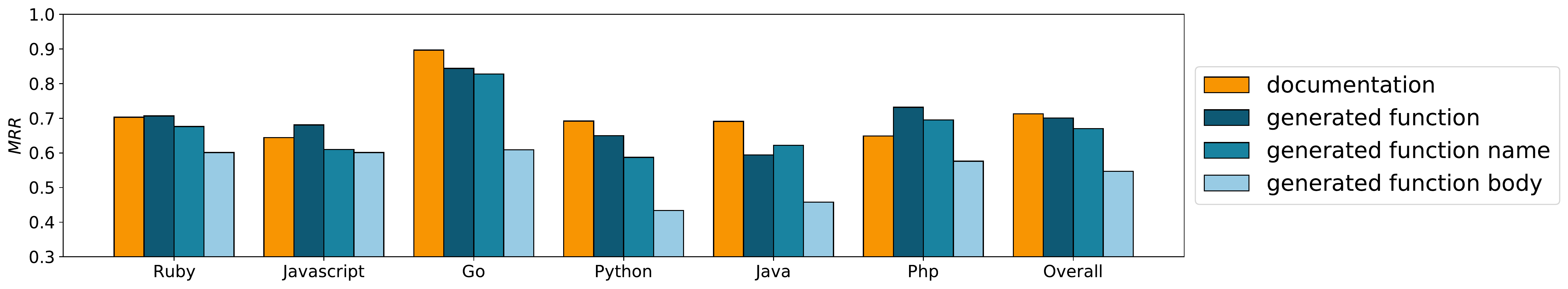}
    \caption{Generated code study on code retrieval task.}
    \label{fig:code_name_body}
\end{figure*}

 \subsubsection{Overall Performance}
 \Cref{tab:main_table_1} shows the main results of all models. GACR is our proposed model, where 'S' represents expanding documentation query with a single generated code snippet (\Cref{subsec:single_code}), 'M' for multiple generated snippets (\Cref{subsec:multi-code}). We highlight the best results for each coding language (each column). 
 
Overall, our proposed models outperform all the baselines, validating the efficacy of the generation-augmented query expansion framework. Even more, GACR-M model exhibits up to $27.8\%$ significant improvement on the \textit{Php} data compared to GraphCodeBERT. 

 \subsubsection{Better Query Quality}
\begin{figure}
    \centering
    \includegraphics[width=1.0\linewidth]{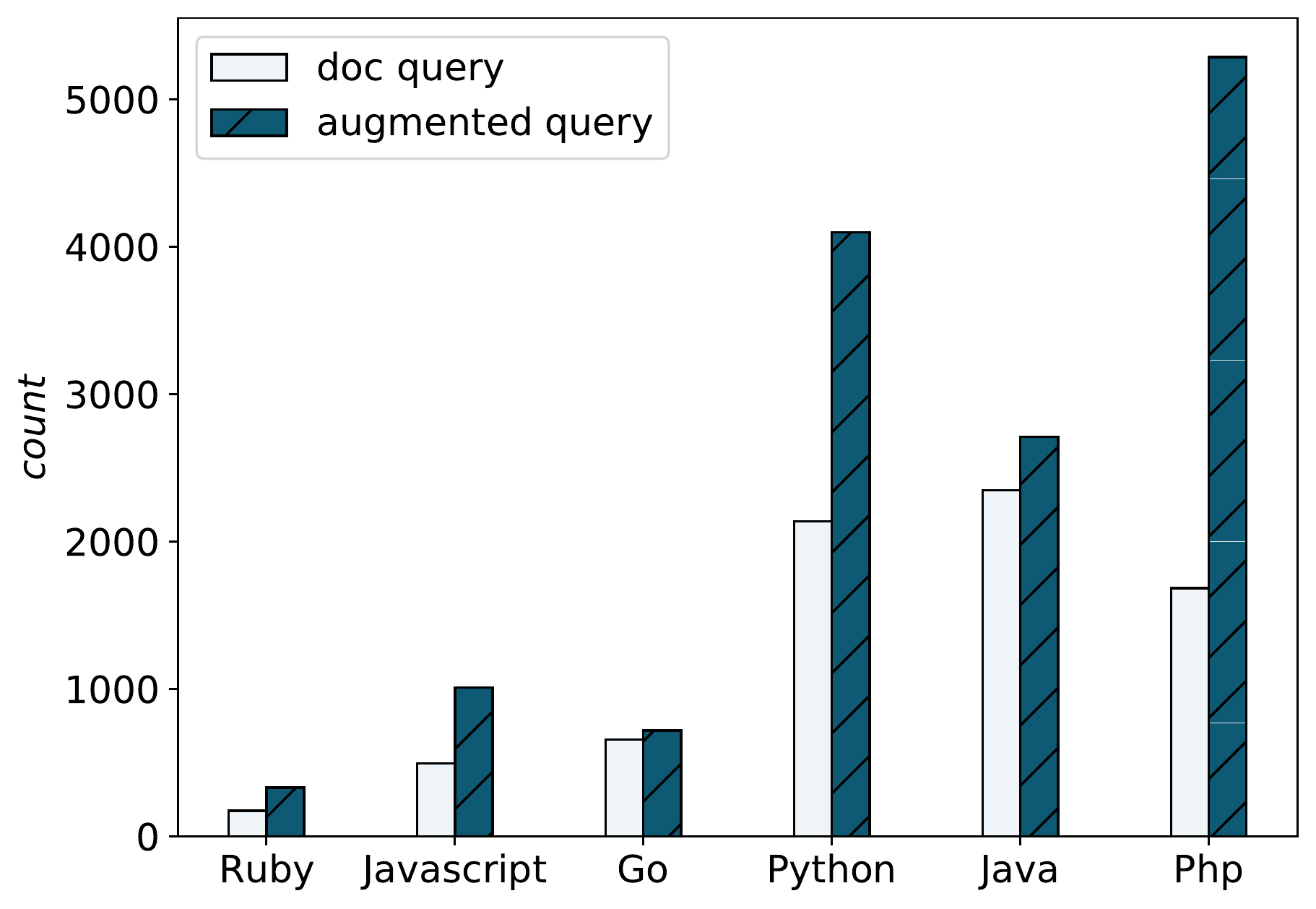}
    \caption{Comparison of GraphCodeBERT (with documentation query) and GACR-S (with augmented query) w.r.t. counting of superior queries.}
    \label{fig:counting}
\end{figure}

\Cref{fig:counting} displays the superiority of the proposed generation-augmented frameworks from the perspective of counting relative ranks. Specifically, we compare the rank of the ground truth code snippet for the same documentation query (with or without augmentation). Then we count how many each has a smaller value (higher rank). From \Cref{fig:counting}, we can observe that augmented query consistently obsess higher rank over all datasets, especially in \textit{php} language. 

 \subsubsection{Pre-train Initialization}
For different pre-training initialization - GraphCodeBERT and CodeRetriever, GACR models consistently get better results while CodeRetreiver-based pre-training is better than GraphCodeBERT-based one, owing to large scale contrastive learning strategy.

 % From the  \Cref{tab:main_table_1}, we have following observations:
 % \begin{itemize}[leftmargin=0.15in]
 % \setlength\itemsep{-0.5em}
 %     \item  In general, our proposed models outperform all the baselines, validating the efficacy of generation-augmented query expansion. Even more, GACR-M-pre model exhibits up to $27.2\%$ significant improvement on the $Php$ data. Overall, both GACR-M-pre and GACR-M-post, can achieve $11\%$ improvement compared to previous state-of-the-art model - GraphCodeBERT.
 %     \item In majority language data, models (GACR-M) that leverage multiple generated code perform better than the one (GACR-S) with single generated code, which demonstrates that more information from different generated code can boost the model further.
 % \end{itemize}

\subsection{RQ2. How could generated codes help retrieval tasks?}

\begin{figure}
    \centering
    \includegraphics[width=0.9\linewidth]{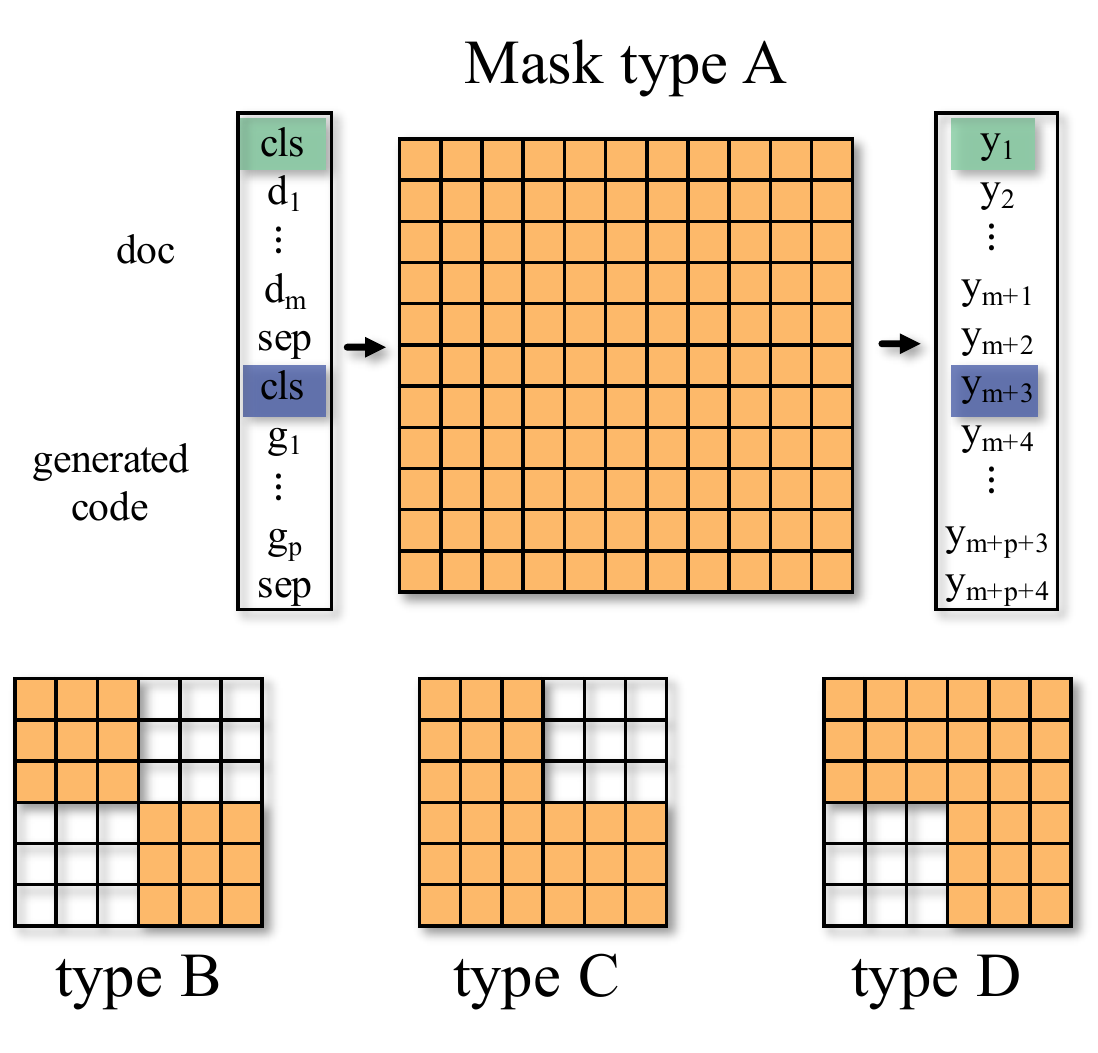}
    \caption{Attention mask paradigms. There A - D are four different types of attention masks, where color graces represent 1 and white represents 0.}
    \label{fig:att4}
\end{figure}

In this subsection, we would like to investigate and answer how the generated code snippets could help the retrieval tasks. We aim to discuss two aspects: the effect of each component in generated code function and a specific case of why generation-augmentation works.
\subsubsection{Function-level generated code performance}

% \begin{figure*}
%     \centering
%     \includegraphics[width = \linewidth]{figure/code_name_body.pdf}
%     \caption{Generated code study on code retrieval task.}
%     \label{fig:code_name_body}
% \end{figure*}

In general, the generated code snippet is presented at a function level. At this point, a code of function can be divided into two parts in both semantic and spatial meanings: the function name (typically the first line of a piece of code, including the input arguments), and the remaining part called the function body. Here we empirically investigate how these different components could help the downstream code retrieval task. We separately treat \textit{documentation}, \textit{generated function} (entire generated code), \textit{generated function name}, \textit{generated function body} as queries to retrieve the code, and the results are presented in \Cref{fig:code_name_body}. For all the cases, we can find that the \textit{generated function name} of the generated code outperforms the \textit{generated function body}, which indicates that typically the name of a function contains sufficient or even more information than the function body. And comparing the \textit{documentation} with \textit{generated function}, sorely treating generated code as a query could achieve comparable performance w.r.t the \textit{documentation} and sometimes even better. Thus, this confirms the benefit to combine these two pieces of information (from generated code and from documentation).
\subsubsection{Case Study}
\begin{figure*}
    \centering
    \includegraphics[width=\linewidth]{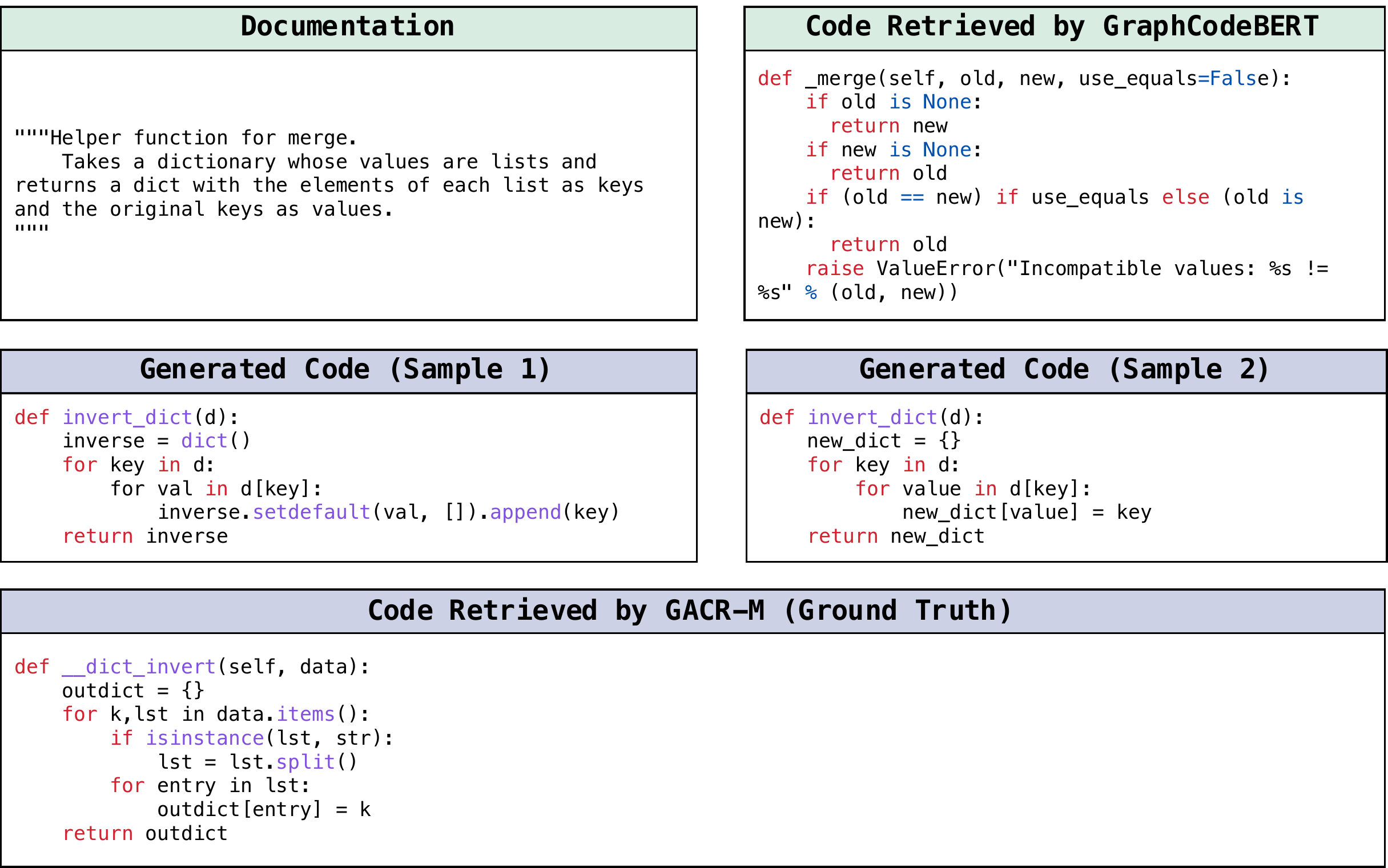}
    \caption{An example in \textit{python} language where generated code boosts the performance of documentation query.}
    \label{fig:python_14840}
\end{figure*}

\Cref{fig:python_14840} gives a specific example of how the generated code helps the documentation query to find the right code snippet. In the base model (GraphCodeBERT), "Documentation" in \Cref{fig:python_14840} serves as a query, aiming to find the "(Ground Truth" while it actually retrieves the "Code Retrieved by GraphCodeBERT". This can be interpreted that without an understanding of the description - "inverse keys and values of a dictionary", the model finds the code snippet by looking up and matching some keywords, for example, "merge", and "values", etc. On the contrary, the generation model could get a better understanding of the description of "Documentation". It indeed creates some semantically meaningful functions ("Generated Code (Sample 1 and 2)"), which share almost the same function name as "(Ground Truth)" as well as the real functionality. By leveraging the power of the generation side, a better interpretation of the documentation leads to more accurate retrieval.

\subsection{RQ3. How effective of each component of the framework (Ablation Study)?}

\subsubsection{Study of the Attention Mechanism}

% \begin{figure*}
%     \centering
%     \includegraphics[width=\linewidth]{figure/slides/4attmask.pdf}
%     \caption{Attention mask paradigms. "doc vec" is the representation vector comes from the "cls" token of the documentation after the encoder while the "gc vec" is the vector from that of generated code. A - D are four different attention masks, where color graces represent 1 and white represent 0.}
%     \label{fig:att4}
% \end{figure*}

In this subsection, we study the affection of attention mechanisms in the proposed framework. Largely, we design four types of masks shown in \Cref{fig:att4} when fusion documentation and code. And the corresponding results are presented in \Cref{tab:nlgc_att}. The Type-A mask is the best overall.

\begin{table}[]
\caption{Effect of Attention Mechanism}
\label{tab:nlgc_att}
\centering
\resizebox{\columnwidth}{!}{%
\begin{tabular}{lcccccccc}
\cmidrule(lr){1-9} 
\multicolumn{2}{c}{Model}                                                                         & \textbf{Ruby}  & \textbf{Javascript} & \textbf{Go}    & \textbf{Python} & \textbf{Java}  & \textbf{Php}   & \textbf{Overall} \\
 \cmidrule(lr){1-9} 
\multicolumn{2}{c}{\small{GraphCodeBERT} }                                                                 & 0.703 & 0.644      & 0.897 & 0.692  & 0.691 & 0.649 & 0.713   \\
 \cmidrule(lr){1-9} 
\multirow{4}{*}{\begin{tabular}[c]{@{}l@{}}GACR-S\end{tabular}} & A & \textbf{0.772} & 0.747      & 0.893 & \textbf{0.768}  & \textbf{0.726} & \textbf{0.813} & \textbf{0.786}   \\
                                                                                              & B & 0.755 & 0.741      & \textbf{0.895} & 0.760  & 0.708 & 0.800 & 0.776   \\
                                                                                              & C & 0.767 & \textbf{0.750}      & 0.865 & 0.763  & 0.720 & 0.811 & 0.779   \\
                                                                                              & D & 0.766 & 0.743      & 0.893 & \textbf{0.768}  & 0.699 & 0.804 & 0.779  \\

 \cmidrule(lr){1-9} 
\end{tabular}
}
\end{table}

% \subsection{Dual Representation}

\subsubsection{Numbers of Augmented Code}
In \Cref{tab:main_table_1}, we observe that GACR-M achieves generally better results than GACR-S, concluding that with more generated code snippets incorporated, the model can get better retrieval results due to more information.
% In \Cref{fig:illustration}, the documentation and multiple generated code snippets are combined before they are fed into the Encoder. 
Due to the limitation of the model, the total length of the input sequence (fusion of documentation and generated code tokens) is limited (typically less than 256). We are able to append more generated code snippets by limiting the maximum length for each code snippet. In \Cref{tab:code_length}, we list three different code length (from $32$ to 128) limits and the results tell that $64$ can be a good balance.

\begin{table}[]
\centering
\caption{Trade-off of augmented code numbers and length limiting, studied on GACR-S model.}
\label{tab:code_length}
\resizebox{\columnwidth}{!}{%
\begin{tabular}{c|ccccccc}
\cmidrule(lr){1-8} 
\textbf{length} & \textbf{Ruby}  & \textbf{Javascript} & \textbf{Go} & \textbf{Python} & \textbf{Java}  & \textbf{Php}   & \textbf{Overall} \\
\cmidrule(lr){1-8} 
32              & 0.786          & 0.761               & 0.894       & 0.769           & \textbf{0.739} & 0.825          & \textbf{0.796}   \\
\cmidrule(lr){1-8} 
64              & \textbf{0.798} & \textbf{0.767}      & 0.898       & 0.766           & 0.686          & \textbf{0.825} & 0.790            \\
\cmidrule(lr){1-8} 
128             & 0.794          & 0.763               & 0.896       & \textbf{0.772}  & 0.679          & 0.813          & 0.786\\
\cmidrule(lr){1-8} 
\end{tabular}
}
\end{table}

\subsubsection{Generation Hidden Representation}
A generation model takes a prompt (NL documentation) as an input and turns it into a hidden representation by an encoder, and further interprets it into the code snippet by a decoder. Instead of utilizing the generated code (like our approaches) to help retrieval, it is also possible to augment the documentation query (in the format of embedding) with intermediate hidden representations (also in the format of embedding vectors) from the generation side. Due to the difficulty of obtaining the intermediate representation from the Codex model, here we conduct the empirical experiments on GPT-J\footnote{\hyperlink{https://huggingface.co/NovelAI/genji-python-6B}{https://huggingface.co/NovelAI/genji-python-6B}} \citep{gpt-j}, a pre-trained model on \textit{python} language.
\Cref{tab:embedding} indicates that the generated code plays a better role in retrieval tasks, which might be owed to the interpretability of the decoder of generation model.

% \usepackage{multirow}
\begin{table}[]
\centering
\caption{Comparision the effects of generated code and hidden representation. (GCB short for GraphCodeBERT, Gen short for generation.)}
\label{tab:embedding}
\resizebox{\columnwidth}{!}{%
\begin{tabular}{|c|c|l|c|}
\hline
Gen             & Retrieval     & Query Format                       & Python \\ \hline
-                      & GCB & documentation                      & 0.692  \\ \hline
\multirow{2}{*}{Codex} & GCB & generated code                     & 0.650  \\ \cline{2-4} 
                       & GACR-S        & augmented with generated code      & 0.768  \\ \hline
\multirow{4}{*}{GPT-J} & GCB & generated embedding                & 0.520  \\ \cline{2-4} 
                       & GCB & generated code                     & 0.566  \\ \cline{2-4} 
                       & GACR-S        & augmented with generated embedding & 0.671  \\ \cline{2-4} 
                       & GACR-S        & augmented with generated code      & 0.758  \\ \hline
\end{tabular}
}
\end{table}

% \begin{table*}[]
% \caption{Comparision the effects of generated code and hidden representation}
% \label{tab:embedding}
% \centering
% \begin{tabular}{ccc}
% \cmidrule(lr){1-3} 
%                                                       & Model                                   & Python \\
%                                                         \cmidrule(lr){1-3}
% \multicolumn{2}{c}{\begin{tabular}[c]{@{}c@{}}Baseline (Doc)\\      GraphCodeBERT\end{tabular}} & 0.692  \\
% \cmidrule(lr){1-3}
% \multirow{2}{*}{Generation Embedding}                 & Embedding as query                      & 0.5204 \\
%                                                       & Doc + embedding as query                & 0.6709 \\
%                         \cmidrule(lr){1-3}
% \multirow{2}{*}{Generation Code}                      & Generated Code as query                 & 0.5664 \\
%                                                       & Doc + GC as query                       &0.7582\\
%                                                         \cmidrule(lr){1-3}
% \end{tabular}
% \end{table*}

\section{Related Work}\label{sec:related}

\subsection{Code Retrieval Tasks}
\textit{Code retrieval} or \textit{code search}, known as to retrieve a relevant code snippet from the candidate pool given a natural language query, has been widely studied for decades spinning from traditional techniques \citep{anne} to deep learning methods \citep{CodeRetriever_Li2022,codebert,guo2021graphcodebert}.
CODEnn \citep{gu2018deepcs} is the early deep learning model that jointly learns embedding vectors for both code snippets and natural language descriptions and further calculates the similarity in the embedding space.
CodeSearchNet \citep{codesearchnet} provides benchmark code search tasks in different programming languages. On top of this benchmark, CodeBERT \citep{condenser} is the first large bimodal
pre-trained model with Masked-Language Modeling (MLM) and replaced token detection objectives.
GraphCodeBERT \citep{guo2021graphcodebert} further improves the performance by utilizing the semantic-level code structure when designing attention pattern. UniXcoder \citep{guo-etal-2022-unixcoder} is a unified  cross-modal pre-trained model which enhance the code representation by leveraging information from code comment and AST. CodeRetriever \citep{CodeRetriever_Li2022} adopts unimodal and bimodal two contrastive learning schemes and achieves the start-of-the-art in code search task. 

% \citep{chai2022cdcs} pre-trained the model in large corpus of common languages and leverages meta-learning on domain-specific languages for code search task.

\subsection{Retrieval and Generation as Augmentation}
\textit{Code retrieval} or other information retrieval models can also serve as an auxiliary unit to help enhance related domain tasks, like code generation, code auto-completion or code summaries, etc \citep{DBLP:journals/ese/XiaBLKHX17}. \citet{hayati-etal-2018-retrieval} propose a model named RECODE to explicitly refer to existing code snippets when generating code, relying on subtree retrieval. \citet{hashimoto2018} propose a retrieve-and-edit framework for code generation and completion by optimizing a joint objective. \citet{retrieval-aug-generation} fine-tune the retrieval-augmented generation (RAG) models end-to-end. REDCODER \citep{parvez2021retrieval} is a retrieval augmentation framework, where it supplies and enhances the code generation or summarization model by searching the relevant codes or summaries from the candidate database. ReACC \citep{lu2022reacc} is a code completion framework that is augmented by leveraging external context, aka retrieving semantically and lexically similar code snippets. GAR \citep{mao-etal-2021-generation} is a Generation-Augmented Retrieval model for open-domain questions answering.
\section{Conclusion}\label{sec:conclustion}

In this paper, we propose a generation-augmented query expansion framework on code retrieval task. To the best of knowledge, this is the first work that leverage generation paradigm to help code retrieval. Specifically, we design patterns that fusion documentation and its generated code function as a expansion query to search related code. We show by cases that generated code from natural language documentation would enhance the semantic similarity to its ground truth code snippet. Extensive empirical experiments validate the achievement of new state-of-the-art results on CodeSearchNet benchmark.

% \section{Future Work}


\bibliography{NLP}
\bibliographystyle{acl_natbib}

\appendix

\section{Case Study}\label{sec:case}
\subsection{Case Study: Augmentation Boosting}
\begin{figure*}
    \centering
    \includegraphics[width=\linewidth]{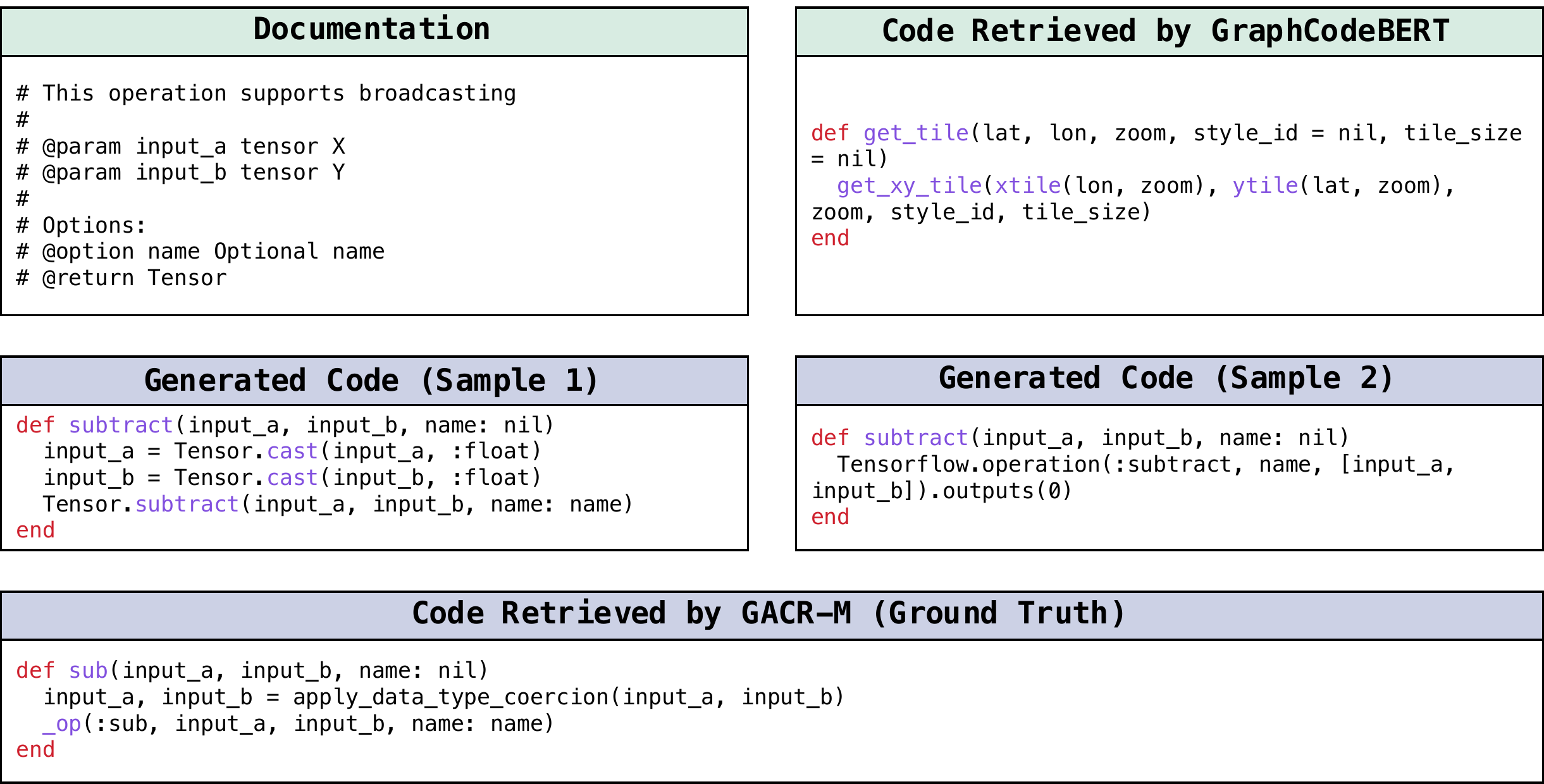}
    \caption{An example in \textit{ruby} language where generated code snippets boost the performance of documentation query.}
    \label{fig:ruby_296}
\end{figure*}

\begin{figure*}
    \centering
    \includegraphics[width=\linewidth]{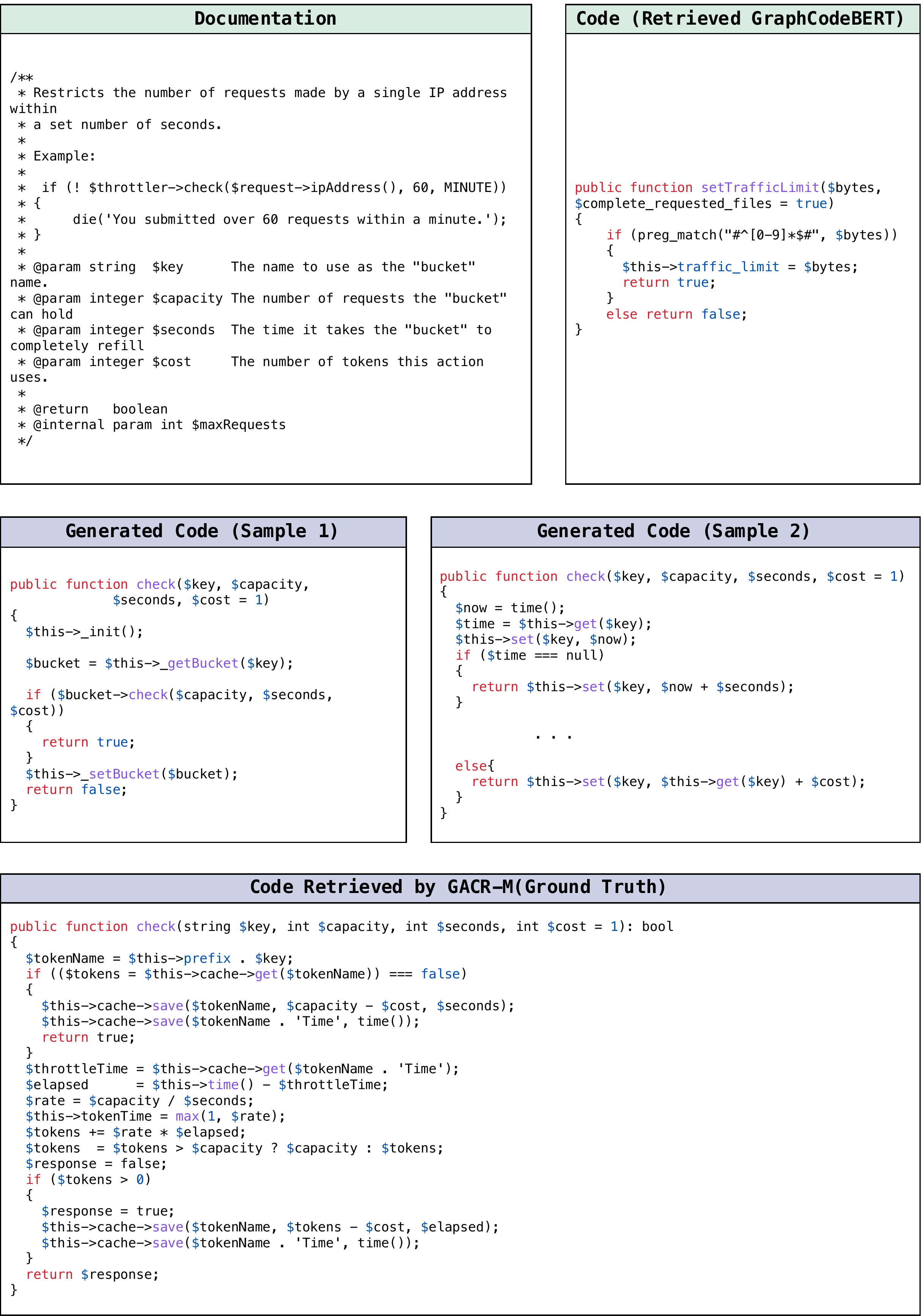}
    \caption{An example in \textit{php} language where generated code snippets boost the performance of documentation query.}
    \label{fig:php_144}
\end{figure*}

In this section, we show some examples (in \Cref{fig:ruby_296,fig:php_144}) that generation-augmented frameworks retrieve better results (or even ground truth code snippets) while models with only documentation queries retrieve unrelated ones.

%\subsection{Ruby}

%\subsection{Javascript}

%\subsection{Python}

% \begin{figure*}
%     \centering
%     \includegraphics[width=\linewidth]{figure/case/python_14840.pdf}
%     \caption{An example in \textit{python} language where generated code snippets boost the performance of documentation query.}
%     \label{fig:python_14840}
% \end{figure*}

%\subsection{Php}

\subsection{Case Study: Augmentation Depressing}

\begin{figure*}
    \centering
    \includegraphics[width=\linewidth]{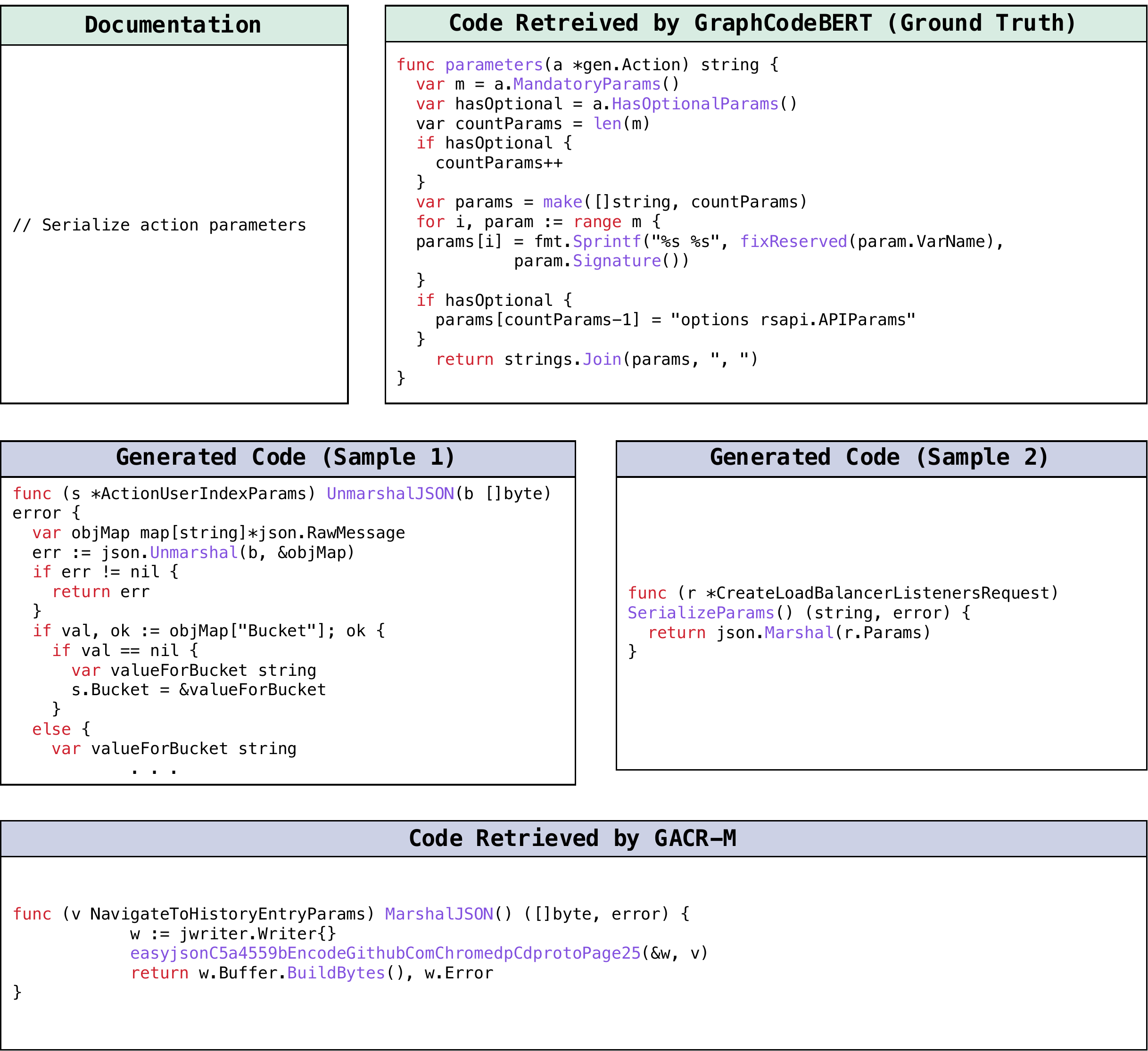}
    \caption{An example in \textit{go} language where generated code snippets depress the performance of documentation query.}
    \label{fig:go_6257}
\end{figure*}

\begin{figure*}
    \centering
    \includegraphics[width=\linewidth]{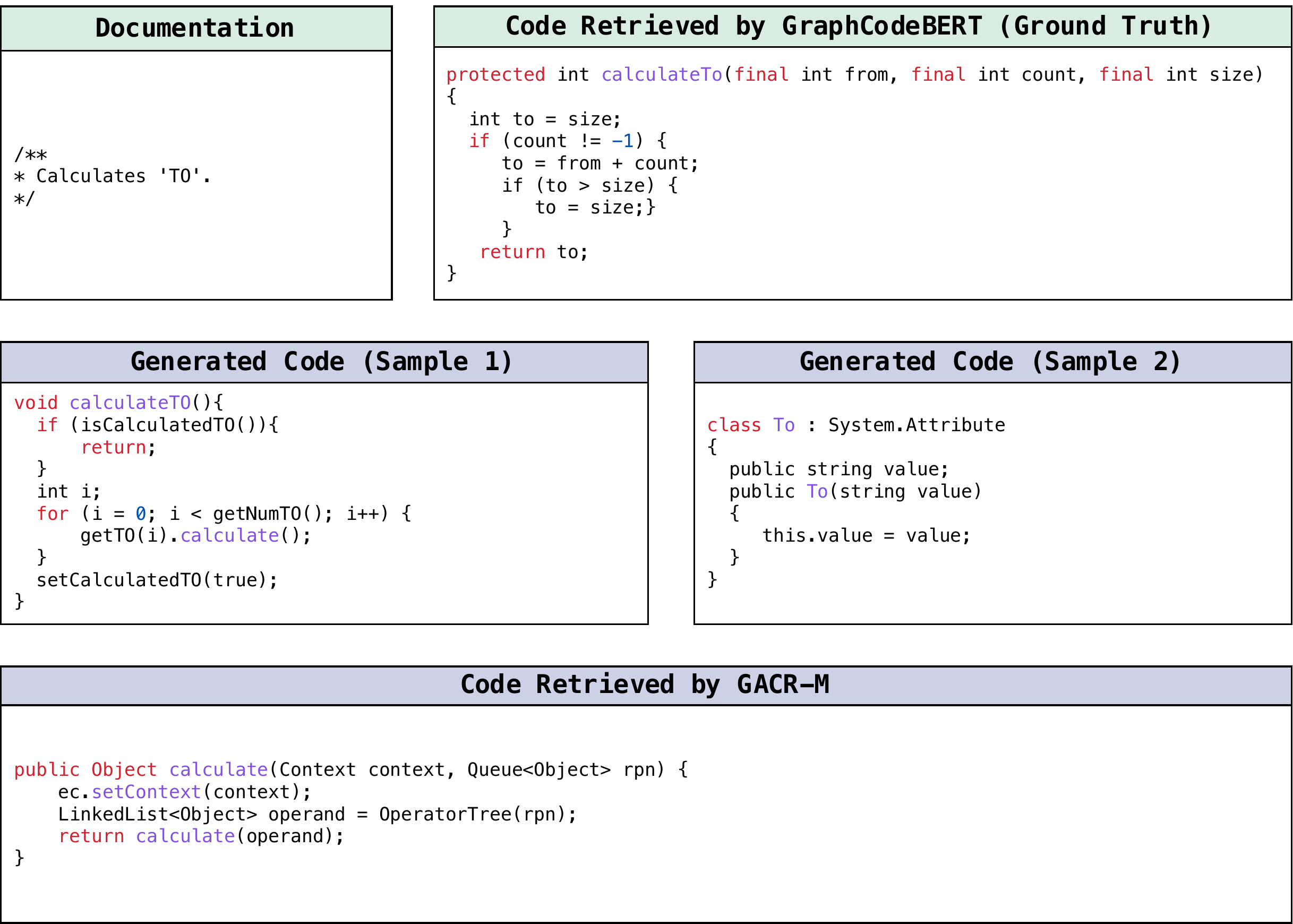}
    \caption{An example in \textit{java} language where generated code snippets depress the performance of documentation query.}
    \label{fig:java_1060}
\end{figure*}

In this section, we show some examples (in \Cref{fig:java_1060,fig:go_6257}) that models with only documentation queries are sufficient to retrieve good results while the generation-augmented frameworks find worse ones.

%\subsection{Go}

%\subsection{Java}

%\subsection{Case Study: Summary}

\end{document}